# Cost Estimation for Alternative Aviation Plans against Potential Radiation Exposure associated with Solar Proton Events for the Airline Industry

Yosuke A. Yamashiki[1], Moe Fujita[2], Tatsuhiko Sato[3], Hiroyuki Maehara[4], Yuta Notsu[5], and Kazunari Shibata[6]

Corresponding to Yosuke A. Yamashiki at

yamashiki.yosuke.3u@kyoto-u.ac.jp

Abstract

In this paper, we present a systematic approach to effectively evaluate potential risk-cost caused by exposure to Solar Proton Events (SPEs) from solar flares for the airline industry. We also evaluate associated health risks from radiation by using ExoKyoto, in order to provide relevant alternative ways to minimize economic loss and opportunity. The estimated radiation dose induced by each SPE for the passengers of each flight is calculated using ExoKyoto and PHITS. We determine a few scenarios for the estimated dose limit at 1 mSv, and 20 mSv, corresponding to the effective dose limit for the general public and occupational exposure, respectively, as well as a higher dose induced by a possible superflare.

[1] Graduate School of Advance of Integrated Studies in Human Survivability, Kyoto University, Sakyo, Kyoto, Japan

[2] Graduate School of Advance of Integrated Studies in Human Survivability, Kyoto University, Sakyo, Kyoto, Japan

[3] Nuclear Science and Engineering Center, Japan Atomic Energy Agency (JAEA)

[4] Okayama Branch Office, Subaru Telescope, National Astronomical Observatory of Japan, NINS, Kamogata, Asakuchi, Okayama, Japan.

[5] Laboratory of Atmospheric and Space Physics, University of Colorado Boulder, Boulder, CO, USA

[6] Unit of the Synergetic Studies for Space, Kyoto University, Sakyo, Kyoto, Japan

We set a hypothetical airline shutdown scenario at 1 mSv for a single flight per passenger, due to legal restrictions under the potential radiation dose. In such a scenario, we calculate the potential loss in direct and opportunity cost under the cancellation of the flight. At the same time, we considered that, even under such a scenario, if the airplane flies at a slightly lower altitude (from 12 km to 9.5 km, corresponding to the slight increase of atmospheric depth from 234 $g/cm^2$ to 365 $g/cm^2$), the total loss becomes much smaller than flight cancellation, and the estimated total dose now goes down from 1.2 mSv to 0.45 mSv, which is below the effective dose limit for the general public.

In the case of flying at an even lower altitude (7 km corresponding atmospheric depth with 484 $g/cm^2$), the estimated total dose becomes much smaller, to 0.12 m Sv. If we assume the increase of fuel cost is proportional to the increase in atmospheric depth, the increase in cost becomes 1.56 and 2.07 for the case of flying at 9.5 km and at 7 km respectively. Lower altitude flights provide more safety for the potential risk of radiation doses induced by severe SPEs. At the same time, since there is total loss caused by flight cancellation, we propose that considering lower flight altitude is the best protection against solar flares.

Introduction

**According to ICAO's Manual on Space Information in Support of International Air Navigation (ICAO 2008), space weather advisory messages must be issued as follows. Solar radiation storms are one type of space weather event that may require a fast response due to the immediacy of their impact. In certain situations, the lead time for a radiation advisory may be only a few minutes at most. In order to avoid radiation exposure, considerations of time, distance, and shielding allow decisive actions for mitigation of the threat. Shielding can mitigate solar radiation storms. Shielding from radiation consists of protection by (a) the overhead atmosphere and by (b) the geomagnetic field. When the field vector is more horizontal than vertical, charged particles are turned aside. The Earth's magnetic field is vertical at the poles and horizontal at the equator, so flying at lower latitudes increases the shielding. As for the FLIGHT CREW, advisories of imminent or on-going disruptions to HF, GNSS, and occurrence of radiation effects allow alternate route planning, or delayed use of polar routes. Options may include: a) Time – Delayed entry into regions specified in the advisory. b) Distance – Not only avoiding specified regions, but in the case of radiation, flying at a non-optimal**

**but lower altitude for more shielding by the atmosphere. The best situation is to be able to plan 12-24 hours ahead of the occurrence, making allowances for flight reroutes, fuel, and crew schedules beforehand. Long-haul flights may be the most problematic as options are constrained by fuel, particularly if the airplane is on route when an unpredicted event occurs. According to ICAO, which defines the current flight operation standards, the policy for avoiding exposure as described above is established. Thus, for people working in the aviation industry, the exposure dose received during working hours is a serious problem.**

The risks of radiation exposure from flying in an aircraft are detailed in various previous studies. Dunster and Mclean (1970) evaluated "man-rad" as $10-25 by classifying a tolerable dose and justifiable dose for professionals and the public. The former is to set the upper limit value of human exposure, and it is decided based on social factors, which is almost equal to the risk in other categories except radiation. They calculate the risk from actual radiation exposure, from the comparison with an age-specific death statistics table from the UK. In principle, it is considered to be somewhat higher than the maximum allowable line L of ICRP. They also mention that the tolerable dose becomes a collective dose, but the details are omitted.

The justifiable dose is determined by economic factors, and because of this, it is not possible to determine numerical values in a general theory, and it is necessary to carry out a cost-benefit balance in the context of a certain situation. The basic concept is not to compare the total costs and total benefits, but rather to compare the marginal cost of dose-reduction with the benefit of reducing the probability of cancer occurrence.

Hedgran (1970) evaluated "man-rad" as $100-250. As a general theory, decision-making in individual human activities or groups in society takes into consideration both cost and benefit. While cost includes direct cost and indirect cost (such as risk, benefit), it is also described as including substantial benefit and moral or ethical imaginary benefit.

For example, if you decide to go somewhere by airplane, Total cost = Airplane cost + Risk when you use the airplane (where, risk = accident rate x Q, Q is implied dollor The equivalent toahuman life) is compared with the benefit of using an airplane.

Otway (1970) evaluated "man-rad" as $200. They stated that risk (or cost)-benefit assessment has come to attract attention in determining the societal acceptability of the technological applications. They conducted extensive reviews on the method of analysis and the amount of life to calculate the amount of man-rad, and then briefly introduced their own special method of estimating cost.

Lederberg (1971) evaluated "man-rad" as $100-250. His calculations are based on the cost of health care and the biological effects of natural radiation in the United States. In other words, he estimates the risk of

radiation exposure, and considers only cancer and mutations at low doses. He estimated that radiation exposure increases the cancer incidence rate by 1% for natural background levels (0.1 rad / year). (300,000 people die annually in the United States from cancer, and 300 of them die by natural radiation.) Despite such calculations, Lederberg is not very optimistic for cost-benefit analysis. One of the reasons is that people prefer qualitative judgment and points out that this cannot be judged quantitatively.

Cohen (1970) evaluated "man-rad" as $250. In his paper, the health and physical aspects of the Plowshare plan are considered, which attempts to use nuclear explosions to facilitate peaceful use, particularly the extraction of gas from gas wells. Since rem is a unit of cost due to radiation exposure in his paper, he proposed a new unit called mer (the opposite of rem) as a unit of benefit that can justify the activity that brought about nuclear explosion.

He did not calculate the amount of merman on his own, but according to the results of the literature survey, it was assumed that $250 would be a reasonable value, which would be $250,000 for a lethal dose of 1000 rem. He stated it does not contradict arbitrage or lifetime earning ability.

Sagan (1972) evaluated "man-rem" as $30. He calculated the total human cost for nuclear power generation at each of the series of processes from uranium mining to spent nuclear fuel reprocessing which is related to nuclear power generation. He calculated it by the figure of radiation exposure and accidental death in a general occupational accident. He employed man-rem $30 as the radiation exposure received by professionals and the general public and the value of life at $300,000 for accidental death in a general occupational accident that is not related to radiation exposure for workers.

Lederberg (BEIR) (1972) evaluated "man-rem" as $12-120. The total cost of healthcare in the United States in 1970 is $8 x $10^{10}$ a year, and the population of the United States is approximately 2 x $10^8$, so the annual cost for healthcare per capita is $400.

Here, considering the genetic effects of radiation, it can be said that the medical cost per person per generation becomes 12,000 (= $ 400 / year × 30 years) dollars.

Next, if we assume the exposure of 5 rem (0.17 rem / year) in 30 years, this increases the amount of illness by 0.5 to 5% (It means that the genetic doubling dose of radiation is 200 to 20 rem), which will increase from 0.1 to 1%, an exposure of 1 rem would increase, accordingly, from 0.1% to 1%, resulting in (12,000 × 0.1 to 1% = $ 12 to $ 120.)

Therefore, if we continue until radiation exposure of 1 rem per one generation for 30 years reaches equilibrium, the amount of illness equivalent to the cost of 12 to 120 dollars per person per 30 years is added. That is, the amount of harm done by 1 rem is equivalent to a cost of $12 to $120 as an integral value over all future generations, regardless of whether or not it reaches equilibrium.

Inaba (1977) proposed that it is important to calculate cost-benefit of radiation exposure not only based on "man-rem" but also based on dollar-value by introducing the announcement of ICRP.

Moreover, guidelines and regulations for radiation exposure to protect passengers and airline companies have recently been issued.

As for EU Member States, in accordance with Council Directive 96/ 29/EURATOM, 13 May,1996, "By ordinance, by May 2000, the necessary measures, such as laws, regulations, and administrative regulations, should be introduced, so that the company will carry out cosmic ray exposure measures for flight crews whose annual exposure dose exceeds 1 mSv in accordance with the directive." This is characteristic of the EU, that many countries take an institutional response.

As for North America and Australia, there are no laws or regulations, but they respond by issuing guidelines etc. independently. For example, in the US:

・"Advises on radiation exposure of aircraft crew" (FAA: Federal Aviation Administration : Advisory Circular 120-52)

· "Advice on training aircraft crews regarding cosmic radiation exposure during operation" (FAA: Federal Aviation Administration: Advisory Circular 120-61)

Australia

・"Recommendation 3022 for Limitation of Ionizing Radiation Exposure" (Australia National Occupational Health and Safety Commission (NOHSC))

・ Developing guidelines for business operators (Australian Radiation Protection and Nuclear Safety Agency: ARPANSA)Etc.

On the other hand, as for Asia, such as Thailand, Indonesia, and Malaysia, no response has been made at the present time. This is because the countries are located at a low latitude where the influence of cosmic rays is small.

In addition, some devices to protect people from radiation exposure in aircrafts have appeared, such as ARMAS Flight Module 5. It measures on-demand, real-time radiation doses from all sources at all altitudes based on the demand of global situational awareness. It has exposure alerts and flight path solutions during significant radiation storms as well as GPS location and Iridium plus Bluetooth connectivity.

In this paper, we propose an insurance design as a measure of cost-benefit analysis and risk management for radiation exposure.

Estimated exposure to ionizing radiation due to secondary generated radioactive particles may cause health risks for passengers taking national and international flights. The exposure level to ionizing radiation induced by solar activities and cosmic rays becomes higher in proportion to the altitude, due to thinner atmospheric depth. The airline industry however, tends to fly at a higher altitude, around 12 km above the ground, situated at the tropopose, which separates the troposphere and stratosphere. Thinner atmospheric depth may reduce fuel consumption for each flight, which is advantageous for economic flights in many situations. Also, since the tropopose is situated above the troposphere, where continuous or sudden vertical convection due to ground heating in meteorological and hydrological processes occurs, flying at a higher altitude than the tropopose provides more stable flights for each airframe, avoiding serious vertical mixing. On the other hand, the reduction of the estimated radiation dose has been raised with increased concern against ionizing radiation exposure, especially for the crew. For general passengers, it is an unnecessary concern for long-term exposure through radiation. However, concern should be raised in the case when extreme radiation exposure is expected upon certain conditions. The most serious scenario happens when Solar Proton Events (SPEs) occur.

The international standards for radiation protection have been studied from the time when radiation use became widespread, such as the application of X-rays to medicine. The International X-ray and Radium Protection Committee (IXRP) was established in 1928 as the organization responsible, and in 1950 the International Commission on Radiological Protection (ICRP) , "ICRP") was reorganized. Since then, the ICRP has issued international recommendations on the basic idea of radiation protection and the standards based on it, and those recommendations are respected in many countries in the world, as well as being applied to laws and safety administrations related to radiation protection. These have been widely adopted. Among the many ICRP functions, the main recommendations, which are the basic ideas regarding radiation protection in general, have been updated almost every 10-15 years. ICRP Publication 60 (also known as the 1990 recommendations) is the most recent. In the ICRP Publication 60, it was recommended that exposures from four natural radiation sources (described later), including aircraft accrual, be included as part of occupational exposure. Thus, we need to consider the effects of radiation exposure for the crew and passengers when flying.

The recommended critical dose for the general public is 1 mSv / year. In most situations, however, we are exposed to a natural dose of around 2.4 mSv/year, so this goal is not easily achieved. There are visible effects due to radiation dose when exposure is over 100 mSv, accordingly a low-level dose in the order of a few mSv is not considered serious or fatal. However, at the same time, the recommended dose to distinguish a radioactive protection area is 1.3 m Sv within three months. According to these trends, to circumvent radiation exposure of more than 1 mSv, it would be a safer criterion for the general public to minimize concern associated to radiation exposure. The potential exposure for an air flight crew and passengers are below a few mSv per year, whereas for astronauts, the average dose per day is a few mSv. In this study, we will therefore consider evaluating the potential risk for a very small level dose for the airline industry.

Methods

Dose evaluation for ionizing radiation for different scenarios have been performed using PHITS introduced into the ExoKyoto program. The outline of the method is written in Yamashiki et al. 2019. The outline is as follows (1) Determine the maximum flare event that occurred in every determined period. (2) Estimate the proton fluence and spectra using one of the hardest spectra as GLE43 (3) Perform Monte Carlo simulation using PHITS for dose calculation in every atmospheric depth to provide value in Gray and Sievert, then (4) Evaluate the final dose by extending the results, applying a filter function to represent magnetosphere. We consider the following five different solar flare scenarios; (1) Flare occurs every 1/10 year (2) Flare occurs every year (Annual Maximum Flare) (3) Flare occurs every 10 years (Decadal Maximum Flare) (4) Maximum flare under current Solar activity (Spot Maximum Flare) and (5) Extreme Theoretical Maximum Flare (Possible Maximum Flare). For each scenario, we calculate the above procedure and estimate the possible radiation exposure during one flare event (over a period of a few hours to one day).

We propose insurance for radiation exposure incidents during airline flights in order to estimate the potential cost for each possible scenario, to circumvent exposure.

Frequency × Severity ＝ Premium

In the above equation, 'Frequency' means a percentage of the incidents that occur, 'Severity' means the amount of damage. 'Other exposure' implies other types of exposure, rather than the target incident. In this study we have not evaluated exposures other than radiation exposure due to secondary particles. 'Premium' is the price of insurance.

**We propose establishment of new regulations to reduce radiation exposure during flights.**
If these new regulations are to be established, and the expected (estimated) dose by radiation exposure exceeds 1 mSv, the flight company should cancel the flight. In this scenario, there may be three options:
1) Cancelling flight for every passenger.
2) Flying at a lower alternative altitude ((a) 9.5 km / (b) 7 km) than the original planned altitude (normal flight altitude)　(12 km).

In the first case (1), the loss will be $25,200 (each flight fee $175 × 144 passengers = Net Sales) which includes opportunity cost.

It also includes the total cost which is equal to 'sales cost' + 'sales, general and administrative expenses' for one flight. We assume a flight from New York to San Francisco. In this case, it will be $11,820, broken down in the following table.

| Type of cost | Amount |
|---|---|
| Fuel cost | $3,000 |
| Personnel expenses for one Airplane | $640 |
| Airport Fee for JFK | $1,089 |
| Airport Fee for SFO | $1,005 |
| Tax | $15.6 |
| Cost of Airplane | $1,783 |
| Personnel Fee at Airport | $4,000 |
| Insurance Fee | $288 |
| Total | $11,820 |

For the second case, if the estimated exposure of radiation exceeds 1mSv, but the flight company will change the flight altitude from 12km to 9.5km, the fuel costs would be multiplied by 1.56, thus, the loss would become $4,680.

For the third case, if radiation exposure exceeds 1mSv but the flight company will change the altitude from 12km to 7km, the fuel costs will be multiplied by 2.07, thus, the loss for them will become $6,210.

**According to the occurrence of Ground Level Enhancement (GLE) in which a neutron particle reaches ground level, we first have to estimate the rough probability of the occurrence of each event, by analyzing historical GLE (from GLE1 to GLE70). Using the current survey we simply estimated the probability of an occurrence of the same scale as GLE42/43 using the X scale. GLE42/43 is considered X13, the probability of which is larger than the GLE 43 event, which is 0.4 (0.25). We calculated the number of GLE incidents, which is larger than GLE 43. We estimated the probability of GLE by using this number and drew a graph with logarithmization. In our calculations, we estimated the probability of a Carrington Event (estimated as X45) as once every 160 years (0.006) since it occurred in 1859, and a Miyake class Event as once every 1300 years (0.0007) because it occurred in 774-775 AD as the magnitude of X1001 (141 times larger than GLE69 with X7.1) .**

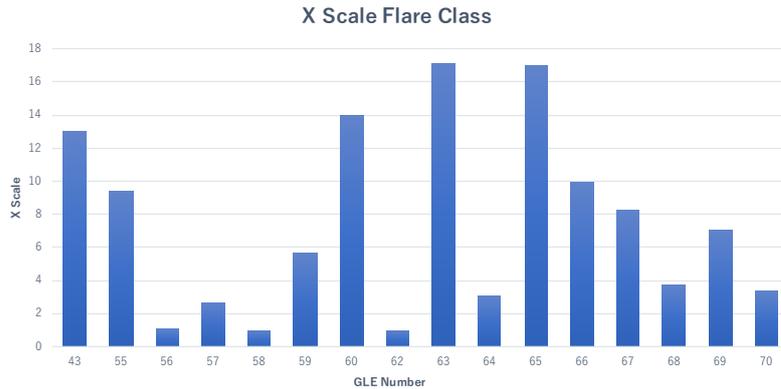

**Figure 1. Major GLE events and their X scale measured by satellite GOES, after GLE 43 occurred in 1989.**

In this study, we estimated the frequency of radiation exposure events, not only accounting for the historical observed GLE events, but also by using the relationship between sunspot area and flare frequency, described in Yamashiki et al. 2019 (Figure 1 Flare frequency vs flare energy for Solar Flares). Using the above, we can estimate the potential global-average radiation exposure at aviation altitude. Please note that this study does not consider the spatial distribution of the location.

**Forecasting Period**

To accurately predict CME that causes severe radiation exposure, we now have to use information captured by the satellite GOES, as well as other means. However, it is difficult to predict. In most cases, the alert will only come a few hours before, when actual optical / X ray signals are captured. For this reason, it is difficult to forecast and make necessary preparations in order to appropriately cancel a flight. According to above analyses, the cancellation of a flight causes huge losses, the alternative of flying at a lower altitude and other similar means will provide a better solution.

**Flight Path**

However, in real flights, a polar flight route will result in a much higher radiation dose compared with lower latitudes, due to the characteristics of energetic particles accumulated in the dypoles of the Earth, known as aurora. Cancellation of flights should be carefully evaluated, considering the flight route for each case. Changing the horizontal flight path to avoid polar regions may be another alternative solution, which should be evaluated in the next study.

Possible Risk for Superflare

As far a flare risks go, what should be considered in the future? According to our scenario, for the very active sun (sunspot area Aspot = 0.003), the theoretical maximum flare energy (Spot Maximum Flare) is calculated at 3.64 x $10^{33}$ ergs. It was already discussed that a superflare may occur for a Sun-like G star according to observations using Kepler (Maehara et al. 2012). This implies we should prepare for many more superflares. In such cases, what would the potential radiation risk be? We set the Possible Maximum Flare, assuming that the sun surface was covered by 20% of sunspots – which is larger than estimated. It should be noted that this value was estimated assuming that sudden atmospheric escape would not happen. If sudden atmospheric stripping occurs the effect would be catastrophic.

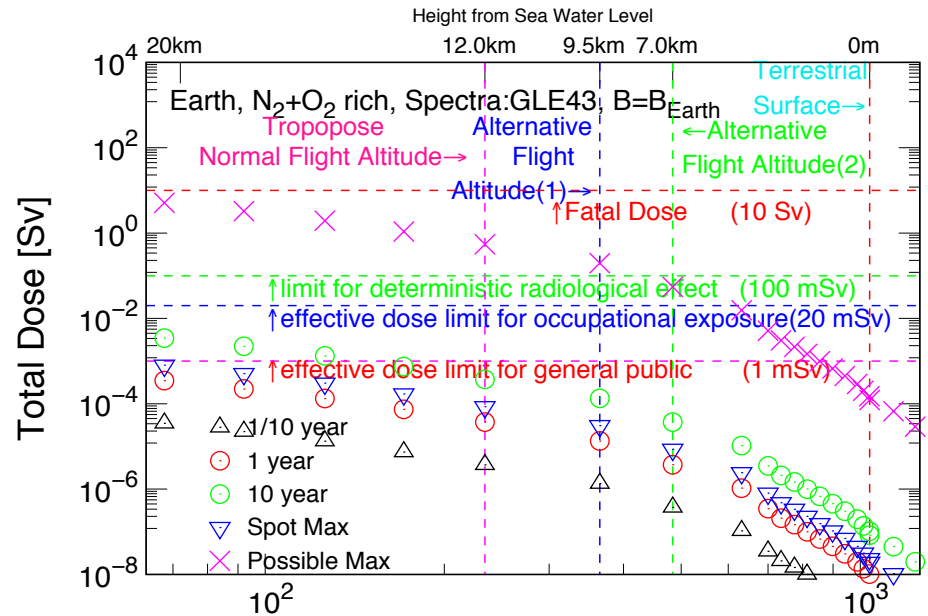

(a) Normal Sun

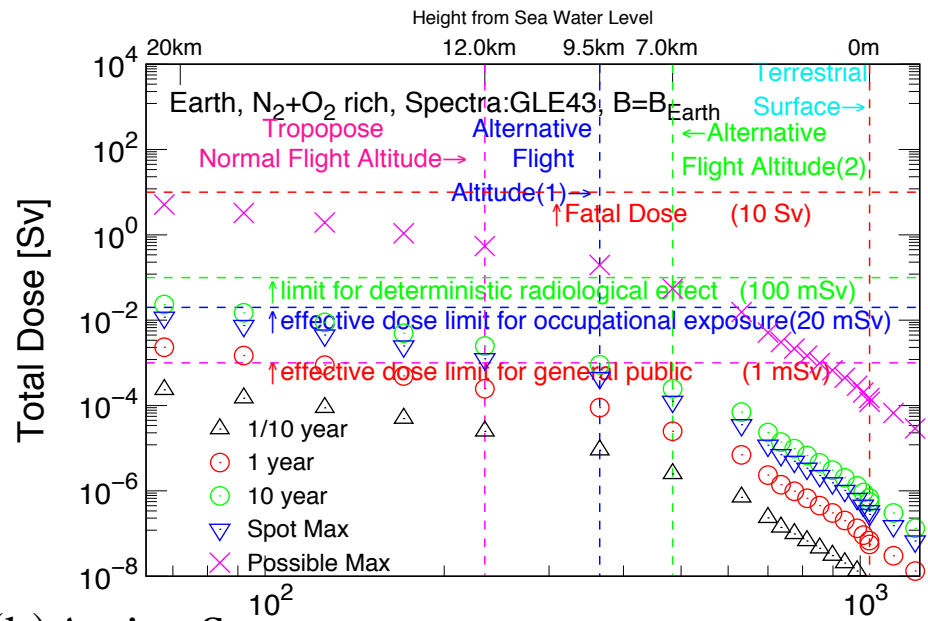

(b) Active Sun

Figure 2 Vertical profile of radiation dose (Sv) on Earth for possible flares on several different scales under normal Sun (normal solar activity period, Aspot = 0.0005) (a) and active Sun ( very active solar period,

Aspot = 0.003), caused by hard proton spectrum (imitating GLE 43) penetrating N2+O2 rich (terrestrial type) atmosphere for Earth with flares every 1/10 year (36 days, black triangle), one year (red circle), decadal (green circle), spot maximum (blue triangle), possible maximum (pink cross). The vertical legend shows the three-typical aviation altitude reference layers: Normal Flight Altitude equivalent to 243 g/cm2, Alternative Flight Altitude equivalent to 365 g/cm2, and Alternative Flight Altitude equivalent to 484 g/cm2, and Terrestrial Surface equivalent to 1037 g/cm2. Note that the value is not identical to the real observation data but the nearest value employed in the Monte-Carlo numerical simulation using PHITS included in ExoKyoto.

**Results and Discussion**

Figure 2 introduces the frequency and energy of Annual Maximum Flares, and Spot Maximum Flares for (a) normal sun (sunspot area = 0.0005) and (b) active sun (sunspot area = 0.003). If we set up different criteria with the figures, we show three critical doses for human life;(i) 1m Sv as the effective dose limit for the general public, determined by ICRP, (ii) 20 mSv as the effective dose limit for occupational exposure, and (iii) **100 mSv,** a limit for deterministic radiological effects. Also, we explicitly displayed 10 Sv as a fatal dose. **According the figures for fatal dose exposure, no scenario offers such a risk, neither for the fatal dose or the deterministic limit, except in the Possible Maximum Flare (PMF) scenario, with the estimated energy as 1.98 X $10^{36}$ ergs.** The PMF is a theoretical maximum flare, which is an occurrence that we normally do not have to consider for our current sun, it would only be considered for extreme conditions. If such a PMF happens, the radiation dose at normal flight altitude (12 km) may reach 0.5 Sv (500 mSv), and becomes much higher than the deterministic dose limit (100 mSv).

We also do not have to consider the exposure risk determined as occupational exposure, as long as flying under the normal altitude.  For these reasons, consideration should be focused only on the exposure for the general public, as 1 mSv per event. Under normal (or quiet) sun with smaller sunspot area (Aspot = 0.0005) no scenario reaches the maximum dose up to 1 mSv at normal flight altitude (12 km) except PMF scenario, whereas under active sun with larger sunspot area (Aspot = 0.003) flares occurs once every 10 years may cause the dose exceed to 1 mSv.

For each of the two cases, the estimated dose may become larger than this limit when flying at the normal flight altitude (12 km), under the Spot Maximum Flare and decadal maximum flare scenarios. In these cases, flying at 9.5 km and 7 km, may provide more safety, while flying at 7 km would be the safest.

Figure 3 introduces the estimated dose in different aviation altitudes by normalized flares. Note that in Figure 2 PMF is the scale of $10^{36}$ ergs, the normalized value reaches a similar value. **It should be noted that since this scale of superflares may never occur on our Sun, this is only considered as a** **theoretical maximum, we do not need to worry about such a radiation dose. However, even with such a dose, the expected dose does not exceed the fatal dose, set at 10 Sv.** Superflares in the scale of $10^{34}$ ergs induce a dose larger than 1 mSv, but not 20 mSv at normal aviation altitude. For a flare with $10^{35}$ ergs the estimated dose becomes larger than 20 mSv, which may be a critical value for the effective dose limit for occupational exposure (20 mSv). In such case, flying at the alternative aviation altitude (1) and (2) gives lower doses than the limit for occupational exposure (20 mSv), but not lower than the limit for the general public (1 mSv). If such a superflare occurs, it might be better to cancel the flight. **However**, it is generally understood that a superflare larger than $10^{34}$ ergs may not be common (it has never been observed in human history) so these extreme conditions are not important for operational levels.

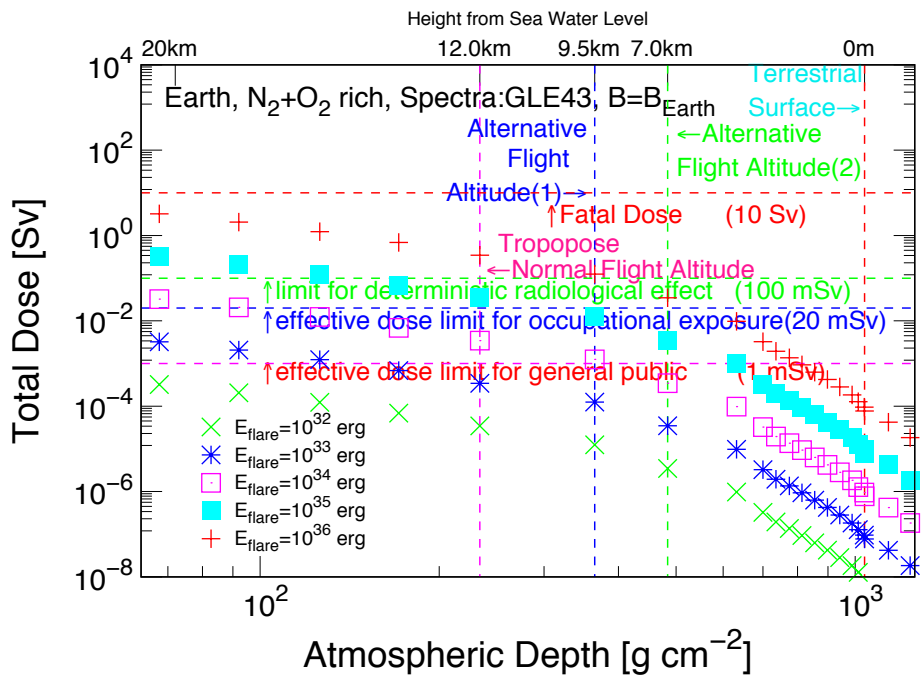

Figure 3 Vertical profile of radiation dose on Earth at different aviation altitudes for normalized flares, caused by hard proton spectrum (imitating GLE 43) penetrating N2+O2 rich (terrestrial type) atmosphere for Earth with $10^{32}$ erg (green cross), $10^{33}$ erg (blue star), $10^{34}$ erg (pink square), $10^{35}$ erg (blue square) and $10^{36}$ erg (red cross) in Sievert (Sv). The vertical legend shows the three-typical aviation altitude reference layers: Normal Flight Altitude equivalent to 243 g/cm², Alternative Flight Altitude equivalent to 365 g/cm², and Alternative Flight Altitude equivalent to 484 g/cm², and Terrestrial Surface equivalent to 1037 g/cm2. Note that the value is not identical to the real observation data but at the nearest value employed in the Monte-Carlo numerical simulation using PHITS included in ExoKyoto.

Conclusion

We evaluated the estimated radiation dose under several SPE scenarios, which may occur either annually or decadally. The estimated dose under the ICRP's expected SPE events did not reach the effective dose limit for occupational exposure (20 mSv) per event. **Accordingly, for these possible scenarios we consider the ICRP's expected SPE events did not reach the effective dose limit for the general public (1 mSv), estimating the cost of legal restriction of an airline.** According to the scenario, we concluded that an alternative flight altitude would provide the most cost-effective alternative solution in such a case, reducing the total exposure of radiation and minimizing the total loss associated with flight changes. Moreover, flying at a lower altitude provides more safety as flying at 7 km will reduce radiation exposure **up to ~ 10%** when compared with flying at 12 km.

**In this study, we only employ simple a case for this evaluation, since in our calculations we only averaged out a global scale estimated radiation dose, without distinguishing the different radiation exposures for different regions of the earth.** Accordingly, in this simulation we could not evaluate how much of a risk may be expected if we fly a polar route, which is normally used for aviation on the Pacific Ocean.

In the next phase of our study, we plan to employ calculations that may provide different doses for different flight paths. **The estimated cost for each radiation exposure scenario in our calculations has not been considered, as the estimated dose is too low to calculate (most of them are below 1 mSv).** At the same time, we may conduct these calculations, assuming the same linear relationship can be applied for radiation exposure below 100 mSv in a future study.

We have not validated individual events and doses observed on airplanes regarding the possibility of the occurrence of superflares in this study. Moreover we should have statistically validated the probability of the occurrence of high SPE and determined the return period for such events. At the same time, it should be noted that we have records of SPEs and Ground Level Enhancement, but we do not have historical records of doses in each SPE. Converting the possibility of flare occurrences into the dose estimation might be the best solution at this stage. For this reason, we consider that our model may provide one of the most effective solutions for aviation exposure event. In a future study, we will be trying to utilize more realistic dose records in each flight record in order to improve our model.

Conflict of Interest: Author has no Conflict of Interest

Research involving human participants and/or animals : This article does not contain any studies with human participations or animals performed by any of the authors.